\documentclass[reprint,english,twocolumn,notitlepage]{revtex4-1}
\usepackage[T1]{fontenc}  \usepackage[latin1]{inputenc}
 \usepackage{babel} \usepackage{txfonts}
 \usepackage{epsfig,epsf,psfrag} 
\usepackage{graphicx} 
\usepackage{subfigure}
\usepackage{pslatex}
\usepackage{epic,eepic} 
\usepackage{color,pstcol}
\usepackage{pstricks}
\usepackage{fancyhdr}  \parindent0em

 \vspace{-10.0cm} 
 
\begin{document} \title{Are thermal fluctuations the sole reason for finite longitudinal resistance in quantum anomalous Hall experiments?} 
  \author{Arjun Mani}
 \author{Colin Benjamin} \email{colin.nano@gmail.com}\affiliation{School of Physical Sciences, National Institute of Science Education \& Research, HBNI, Jatni-752050, India}
\begin{abstract}
In some recent experiments [A. J. Bestwick, et. al., Phys. Rev. Lett. 114, 187201 (2015), Cui-Zu Chang, et. al., Nat. Materials. 14, 473-477 (2015)] it has been shown that in observations of the quantum anomalous Hall (QAH) effect the longitudinal resistance $R_L$ increases as temperature $T$ increases, while Hall resistance $R_H$ loses its quantization with increase in $T$. This behavior was explained due to increased thermal fluctuations as $T$ increases. We show that similar effects arise in QAH samples with quasi-helical edge modes as disorder increases in presence or absence of inelastic scattering even at temperature $T=0$.
 \end{abstract}
\maketitle
\section{Introduction} The experimental realization of quantum Hall(QH) effect in a 2DEG in presence of a magnetic field opened a new direction in dissipation less transport enabled via edge modes\cite{klitzing}. These chiral topological edge modes are potentially useful in low power information processing \cite{roth}. The recent discovery of another cousin brother quantum anomalous Hall (QAH) effect\cite{cui, kou, che} opens up the possibility of its use as current carrying dissipation less edge modes even in absence of magnetic field. In addition, due to the spin polarization of these QAH edge modes, they can be used in spintronic devices as well. The physical origin of QAH effect is different from that of the QH effect, while QH effect was observed in a 2DEG in presence of a magnetic field, QAH effect is observed in ferromagnetic topological insulators in presence of intrinsic spin orbit coupling\cite{wang, wang1}. QAH edge modes consist of chiral propagation of electrons with complete spin polarization, i.e., at one edge of a 2D sample say, spin up/down electron is moving to the right then and at the other edge of the sample spin up/down electron is moving to the left. The longitudinal resistance is zero in case of a single chiral(topological) QAH edge mode. However, experiments\cite{kou,che,wang1} reported a finite longitudinal resistance. These experiments explained their results by speculating that a QAH chiral edge mode always appears in conjuction with quasi-helical QSH edge modes, see Refs. \cite{kou, che, wang1, wang}. This is reasonable since the QAH edge modes seen in Refs.~\cite{kou, che, wang1, wang} are generated from helical QSH edge modes by applying an extra ferromagnetic layer and opening a gap between the pair of helical edge modes. In this way, by splitting the helical edge modes and then suppressing one of them, a single chiral QAH edge mode is sought to be generated. The experimental results of \cite{kou,che,wang1} and the subsequent  interpretation\cite{wang} of these experiments  as seeing not just a chiral(topological) QAH edge mode but in addition also a pair of quasi-helical QSH edge modes were the only game in town until the experiments of Refs.\cite{bestwick,cui}. In Refs.~\cite{bestwick, cui}, the conductivity of a six terminal QAH bar at finite temperature is probed. It is shown in these works that the longitudinal conductivity $\sigma_{xx}$ decreases with increasing reciprocal temperature ($1/T$). The longitudinal resistance $R_{L}$ and Hall resistance $R_H$ are related to the longitudinal conductivity $\sigma_{xx}$ and Hall conductivity $\sigma_{xy}$ as follows, (see page 9-12 of Ref.~\cite{dong})-
\begin{equation}  
R_{L}=\frac{\sigma_{xx}}{\sigma_{xx}^2+\sigma_{xy}^2},\qquad  R_{H}=\frac{\sigma_{xy}}{\sigma_{xx}^2+\sigma_{xy}^2}.
\end{equation}
 When we convert the longitudinal conductivity of \cite{bestwick} to longitudinal resistance via the relation (Eq. (1)), we see that the longitudinal resistance increases with increasing temperature, as in Fig.~1(a). Similarly, if we convert the Hall conductivity of Ref.~\cite{bestwick} to the Hall resistance, we see that it loses its quantization as shown in Fig.~1(b).
\begin{figure}
\centering
{\includegraphics[width=0.5\textwidth]{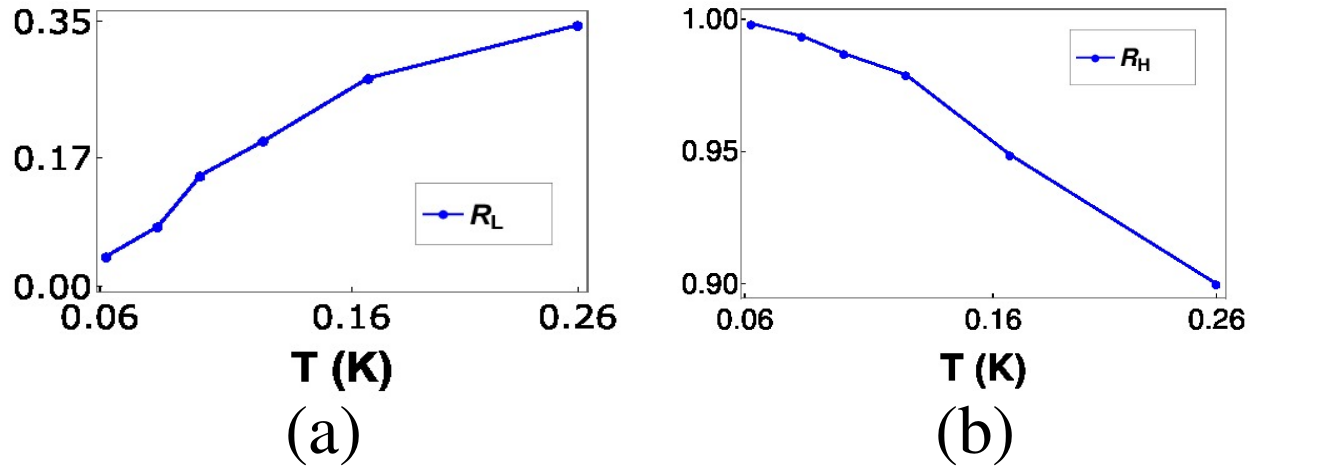}}
 \caption{(a) The longitudinal resistance and (b) the Hall resistance (both in units of $e^2/h$) as in Ref.~\cite{bestwick} for a six terminal QAH sample, adapted from Figure 3 of Ref.~\cite{bestwick}.}
\end{figure}

\begin{figure*}\label{fig2}
  \centering    { \includegraphics[width=.85\textwidth]{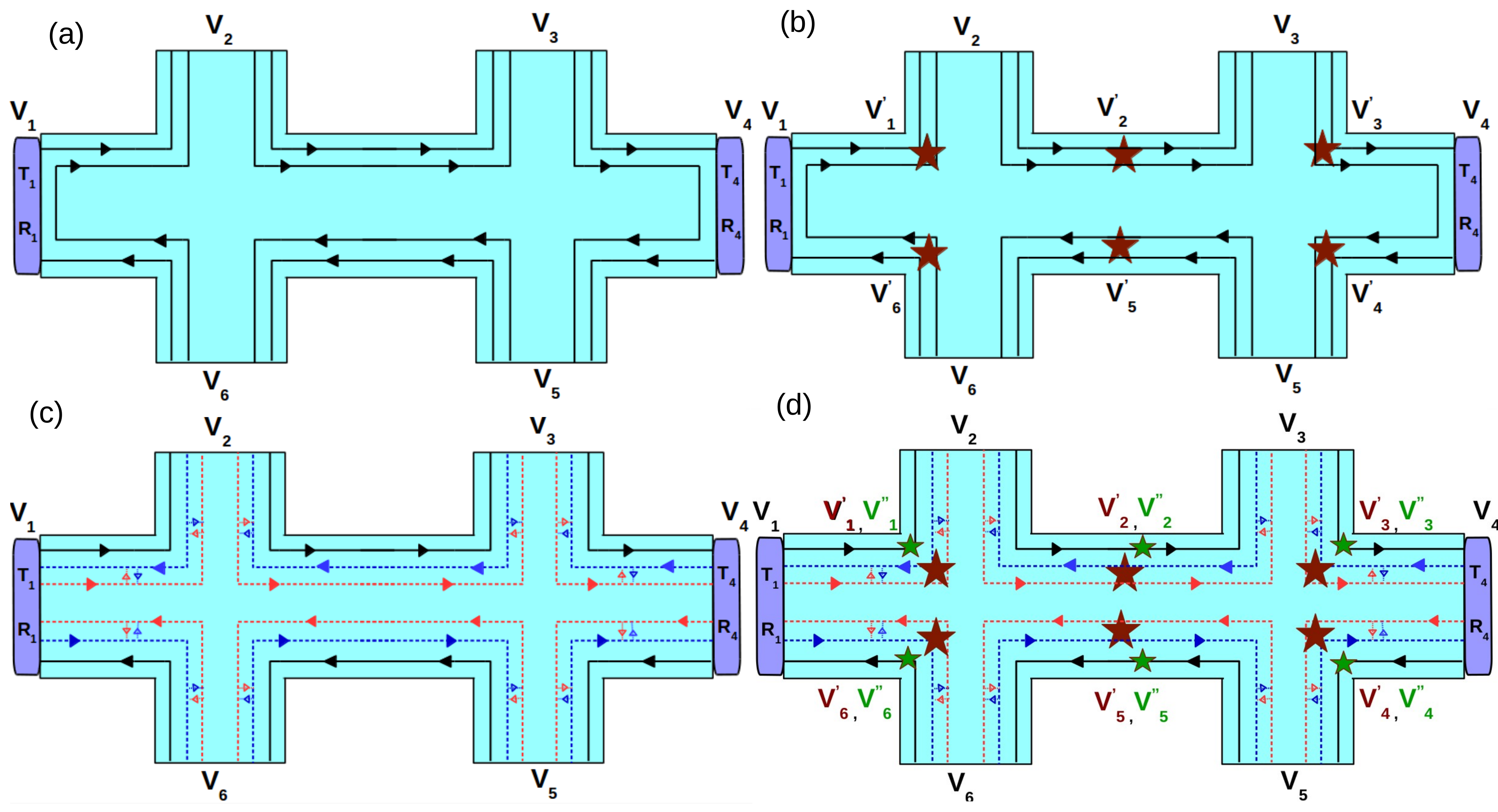}}\caption{(a) A single chiral topological QAH edge mode with disordered contacts $1$, $4$. {Here the quantum anomalous Hall bar only has one edge mode. To show that edge modes are partially reflected and partially transmitted at the disordered contacts and their trajectory from one contact to another after reflection, we have drawn two lines, although both relate to the same edge mode}, (b) single chiral topological QAH edge mode with inelastic scattering and disordered contacts $1$, $4$, (c) single chiral topological QAH edge mode in conjunction with quasi-helical QSH edge modes with disordered contacts $1$, $4$, (d) single chiral topological QAH edge mode in conjunction with quasi-helical QSH edge modes and inelastic scattering with disordered contacts $1$, $4$. Black solid line represents a spin up chiral topological QAH edge mode while red and blue dotted lines represent spin up and spin down quasi-helical QSH edge modes respectively. Green and brown starry blobs on the edge modes imply the equilibration of the energy and population of the chiral topological QAH edge mode and quasi-helical QSH edge modes respectively.}
\end{figure*} 
{ Similar characteristics of Hall and longitudinal resistances with temperature have also been observed in Refs.~\cite{mogi, liu}. In Ref.~\cite{mogi} it's been shown that for temperatures below 1K $R_L$ is zero and $R_H$ is quantized, but above that temperature $R_L$ increases exponentially while $R_H$ loses its quantization. It has been explained therein that this finite $R_L$ and non-quantized $R_H$ are due to inhomogeneous coupling of Dirac electrons and magnetic impurities in the topological insulator and the presence of  bulk conduction along with edge conduction. In Ref.~\cite{liu} on the other hand, finite $R_L$ has been explained as due to the change in transmission probability via domain walls in the topological insulator wherein the transmission probability becomes temperature dependent.} In this letter we theoretically scrutinize a six terminal QAH bar, focusing on the longitudinal resistance $R_{L}$ and Hall resistance $R_{H}$ at zero temperature, see Fig. 2. We have followed the Landauer-Buttiker formalism to calculate the resistances at zero temperature. We find that the longitudinal resistance for  chiral (topological) QAH edge mode with quasi-helical QSH edge modes increases with disorder, while the Hall resistance decreases, which is similar to the behavior observed as function of temperature. { Thus our aim in this paper is to seek to find whether disorder on its own or aided by inelastic scattering can explain the finite longitudinal resistance and loss of quantization of Hall resistance at zero temperature.} We conclude that experiments like Refs.~\cite{bestwick, cui} can not be just interpreted as indication of chiral QAH effect but rather of QAH edge mode occurring with quasi-helical QSH edge modes too at zero temperature. Further temperature is not the sole reason for seeing finite $R_L$, existence of quasi-helical edge modes could also be a plausible reason.

The rest of the paper is organized as follows- first we discuss the Landauer-Buttiker formalism at zero temperature, required to calculate the longitudinal and Hall conductivity in a six terminal QAH sample. We then discuss two cases- 1) a single chiral (topological) QAH edge mode existing in a six terminal QAH sample, and 2) a single chiral (topological) QAH edge mode occurring in conjuction with quasi-helical QSH edge modes in a six terminal QAH sample, from now addressed as QAH+, to distinguish it from the previous case. In each case we calculate the longitudinal resistance and Hall resistance for a) disordered sample and b) disordered sample with inelastic scattering at zero temperature. We bring out the fact that the model of a chiral QAH edge mode existing with quasi-helical QSH edge modes can explain the results of Refs.~\cite{bestwick, cui} at $T=0$. We end the manuscript with some concluding remarks.

\subsection{Landauer-Buttiker formalism} The Landauer-Buttiker formula relating currents ($I_i$) and voltages ($V_i$) at zero temperature \cite{buti, nikolajsen, arjun} for multi-terminal devices is defined as-
\begin{eqnarray}
I_i&=&\frac{e^2}{h}\sum_{j\neq i,\sigma, \sigma'=\uparrow/\downarrow}(T^{\sigma\sigma'}_{ji} V_i-T^{\sigma\sigma'}_{ij} V_j),\\
\text{with }T^{\sigma\sigma'}_{ij}&=&Tr[s^{\sigma\sigma'\dagger}_{ij}s^{\sigma\sigma'}_{ij}] \text{ when $i\neq j$ or $\sigma\neq\sigma'$}\\
\text{and } T^{\sigma\sigma}_{ii}&=&(N_{\sigma}-Tr[s^{\sigma\sigma\dagger}_{ii}s^{\sigma\sigma}_{ii}]),
\end{eqnarray}
where $e$ is the electronic charge, $h$ is Planck constant, $T^{\sigma\sigma'}_{ij}$ is the transmission probability from terminal $j$ to terminal $i$ with initial spin $\sigma'$ to final spin $\sigma$ (where $\sigma,\sigma'=\uparrow/\downarrow$) of electrons, $N_\sigma$ is the number of edge modes with spin $\sigma=\uparrow/\downarrow$, $s_{ij}$ is the element of the scattering matrix $S$ which relates the incoming edge mode amplitudes with the outgoing, and $V_i$ is the potential bias applied at terminal $i$ with $i,j=1-6$ (in our case). 

\section{Disordered 6T QAH bar}
A  six terminal Hall bar with a single chiral (topological) QAH edge mode is shown in Fig.~2(a). 
Herein, we analyze the longitudinal and Hall resistances at zero temperature. Two of the contacts $1$ and $4$ are disordered, i.e., transmission probability $T_i(=1-D_i)$ through those contacts (`$i$' is the index of disordered contact) are less than unity. $D_i(=R_i$, reflection probability at contact `$i$') implies the strength of disorder at contact $i$, which ranges from $0<D_i<1$. $D_i=1$ implies the contact to be completely disordered and $D_i=0$ implies the contact to be ideal (without disorder). The transmission probabilities, $T^{\sigma\sigma'}_{ij}$'s can be calculated following Eqs.~(3,4), where $N_\sigma$ is the no. of edge modes carrying electrons of spin $\sigma$ at contact $i$ and $s_{ij}$ is the element of scattering matrix $S$ (given in the Appendix section A). The transmission probability for an electron from contact `$6$' to contact `$1$` is $T_{16}=\sum_{\sigma\sigma'}T^{\sigma\sigma'}_{16}=T_1=(1-D_1)$ (the transmission probability for a spin up electron from contact $6$ to contact $1$ without a spin flip is $T_1$, i.e., $T^{\uparrow\uparrow}_{16}=T_1$, while there is no transmission of spin down electron, so $T^{\uparrow\downarrow}_{16}=T^{\downarrow\uparrow}_{16}=T^{\downarrow\downarrow}_{16}=0$, see Fig.~2(a)). The  relation between  currents and voltages at  various contacts are deduced following the Landauer-Buttiker formalism (see Eq.~(2)):
\begin{equation}
I=GV,
\end{equation}
where, $I=(I_1,I_2,I_3,I_4,I_5,I_6)^T$ and $V=(V_1,V_2,V_3,V_4,V_5,V_6)^T$ with the conductance matrix as shown below-
\begin{equation}
G=\frac{e^{2} }{h} \left( \begin{array}{cccccc}
    T_1  & 0 & 0 & 0&0&-T_1 \\
    -T_1 & 1 & 0 & 0&0&-R_1\\ 
   0  & -1 & 1& 0&0&0 \\
     0  & 0 & -T_4&  T_4&0&0 \\
      0  & 0 & -R_4 &  -T_4&1&0 \\
      0  & 0 & 0&  0&-1&1 \\
      \end{array} \right).
\end{equation}
Choosing reference potential $V_{4}=0$ and $I_2=I_3=I_5=I_6=0$ (as these are voltage probes) we get the relations between $V_i$'s like $V_2=V_3=T_1V_1$ and $V_5=V_6=V_4=0$. So the longitudinal resistance $R_L^{QAH}=(V_2-V_3)/I_1=0$ and Hall resistance $R_H^{QAH}=\frac{h}{e^2}$. Thus, for a single chiral QAH edge mode the longitudinal and Hall resistance do not get affected by the presence of disorder at contacts $1$ and $4$.

\subsubsection{All probe disorder}
Herein, we calculate the longitudinal and Hall resistance when all contacts are disordered. For all disordered contacts there will be multiple paths from one contact to another. An electron after emerging from one contact, say contact 1 can reach another contact, say contact 3 directly after reflecting at contact 2 with scattering amplitude $t_1r_2t_3$, but this is only one of the many paths. It can also reflect at contact 3 and then at contacts 4, 5, 6, 1, 2 and then finally enters contact 3 with scattering amplitude $t_1r_2t_3(r_1r_2r_3r_4r_5r_6)$. Summing all these paths we get the total scattering amplitude for an electron from contact 1 to 3 as $s_{13}=t_1r_2t_3/a$, where $a=1-r_1r_2r_3r_4r_5r_6$. Thus the transmission probability $T_{13}==T_1T_3R_2/a^2$. Similarly, all other transmission probabilities $T_{ij}$'s for $i,j=1-6$ can be calculated. The current-voltage relations derived from scattering matrices and following Eqs.~(2, 3, 4) are-
\begin{eqnarray}
I=GV
\end{eqnarray}
with
\begin{equation}
G=\frac{e^{2} }{h} \left( \begin{array}{cccccc}
    T_{11}  & -T_{12} & -T_{13} & -T_{14}  & -T_{15} & -T_{16}\\
    -T_{21}  & T_{22} & -T_{23} & -T_{24}& -T_{25} & -T_{26} \\ 
    -T_{31}  & -T_{32} & T_{33} & -T_{34}& -T_{35} & -T_{36} \\
     -T_{41}  & -T_{42} & -T_{43} & T_{44} & -T_{45} & -T_{46}\\ 
     -T_{51}  & -T_{52} & -T_{53} & -T_{54}& T_{55} & -T_{56} \\
     -T_{61}  & -T_{62} & -T_{63} & -T_{64} & -T_{65} & T_{66}\\ 
     \end{array} \right),
\end{equation}
Choosing reference potential $V_{4}=0$ and $I_2=I_3=I_5=I_6=0$ (as these are voltage probes) we get the relations between $V_i$'s, we have longitudinal resistance $R_L^{QAH}=(V_2-V_3)/I_1=0$. Thus, we see even if all the contacts are disordered, the longitudinal resistance still vanishes, while the Hall resistance slightly deviates from its quantized value.

\subsection{Disordered 6T QAH bar with inelastic scattering}
Herein we analyze the longitudinal and Hall resistances in presence of both disorder as well as  inelastic scattering and as before, contacts ($1,4$) are disordered as shown in Fig.~2(b). The probability of an electron to get reflected or transmitted through a disordered contact is $R_i(=D_i)$ and $T_i(=1-D_i)$ respectively with $T_i+R_i=1$ and $i=1,4$. If the length between two contacts is larger than the inelastic scattering length, then edge modes are inelastically scattered which is shown by the brown colored starry blobs in Fig.~2(b). Inelastic scattering equilibrates the population and energy of the electrons coming from two or more different contacts via electron-electron interactions {at zero temperature} or electron-phonon interactions {at finite temperature}. {Inelastic scattering considered in our paper via equilibration of energy of the edge modes is different to the widely used Buttiker's voltage probe model. Inelastic scattering via Buttiker's voltage probe refers to inelatic scattering happening at a particular place between two contacts\cite{yan, rok}, while in our case it can happen any where throughout the region between those contacts\cite{buti, jen}. However, both the models leads to same result.} The edge modes existing between contacts $1$ and $2$ are equilibrated to a new potential $V'_1$, similarly the edge modes occurring between contact $V_i$ and $V_{i+1}$ are equilibrated to $V_i'$, where $i=2-5$ and between contacts $6$ and $1$ to $V_6'$. As the current contacts $1$ and $4$ are disordered, the transmission probabilities through these contacts are $T_1$, $T_4$. Thus, currents and voltages at contacts 1 to 6 are related by the following equations-
\begin{eqnarray}
I_1&=&\frac{e^2}{h}T_1(V_1-V'_{6}),  \qquad I_4=-I_1, \nonumber\\
I_i&=&\frac{e^2}{h}(V_i-V'_{i-1})\qquad{\text{for $i= 2,3,5,6$}}.
\end{eqnarray}
Choosing the reference potential $V_{4}=0$ and $I_2=I_3=I_5=I_6=0$ (since these are voltage probes), we get $V_{2}=V'_1$, $V_3=V_2'$, $V_5=V_4'$ and $V_6=V_5'$. Equilibration (say at $V'_1$) can be explained as follows- the total current coming into $V'_1$ consists of -(i) an edge mode, originating at $V'_6$ with current $\frac{e^2}{h}R_1V'_6$, and (ii) an edge mode coming from $V_1$ with current $\frac{e^2}{h}T_1V_1$. The sum of these two incoming currents $\frac{e^2}{h}R_1V'_6+\frac{e^2}{h}T_1V_1$ is equilibrated to the outgoing current $\frac{e^2}{h}V'_1$ via the potential $V'_1$. Similarly, equilibration at other starry blobs can be explained. The equations resulting from the equilibration's are given below-
\begin{eqnarray} 
\frac{e^2}{h}R_1V'_6+\frac{e^2}{h}T_1V_1&=\frac{e^2}{h}V'_1,\quad \frac{e^2}{h}V_6=\frac{e^2}{h}V'_6,\quad \frac{e^2}{h}V'_5=\frac{e^2}{h}V_5, \nonumber\\
\frac{e^2}{h}R_4V'_3+\frac{e^2}{h}T_4V_4&=\frac{e^2}{h}V'_4,\quad \frac{e^2}{h}V_3=\frac{e^2}{h}V'_3,\quad \frac{e^2}{h}V'_2=\frac{e^2}{h}V_2. 
\end{eqnarray}
From Eq.~(8) and Eq.~(7) we get the relations between the contact and equilibrated potentials as- $V_2=V_3=V_1'=V_2'=V_3'$ and $V_5=V_6=V_6'=V_5'$, which gives the longitudinal resistance  $R_{L}^{QAH}=R_{23,14}=0$ and the Hall resistance $R_{H}^{QAH}=R_{26,14}=\frac{h}{e^2}$. We see that for QAH (topological) case the longitudinal and Hall resistances are independent of inelastic scattering and disorder at zero temperature and are quantized. Next, we consider a Hall bar with a chiral QAH edge mode existing with quasi-helical QSH edge modes, or the QAH+ case.
\begin{figure}
\centering
{\includegraphics[width=0.5\textwidth]{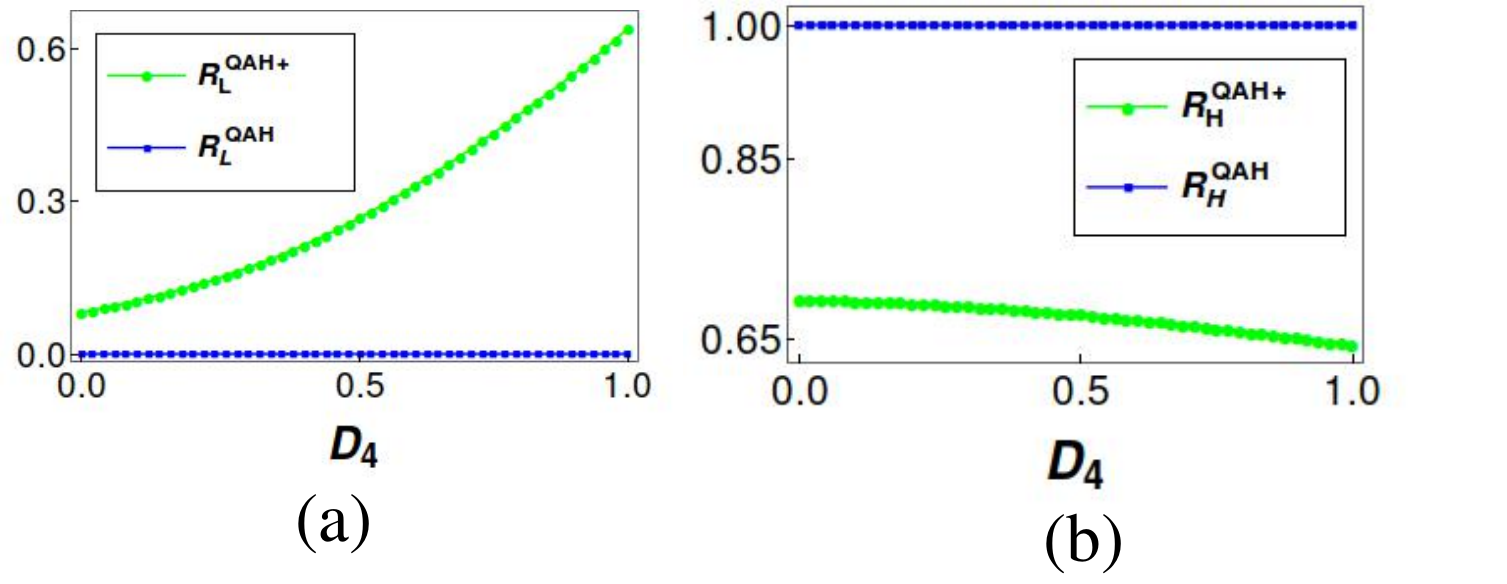}}
 \caption{(a) The longitudinal resistance and (b) the Hall resistance (both in units of $e^2/h$) vs disorder ($D_4$) at contact $4$ is shown for a 6 terminal QAH sample at $T=0$K.  The ``QAH+'' in the superscript of resistance implies that the chiral topological QAH edge mode exists in conjunction with quasi-helical QSH edge modes while `QAH' implies a single chiral QAH edge mode alone. Parameters are $D_1=0.5$, spin-flip scattering $f=0.9$.}
\end{figure}

\section{Disordered 6T QAH+ bar} The six terminal QAH bar is shown in Fig.~2(c). In addition to the QAH edge mode (spin-up polarized) shown as a black solid line, there are two quasi-helical edge modes shown as red (spin-up) and blue (spin-down) dotted lines. Since these quasi-helical helical edge modes are prone to spin flip scattering. We introduce `$f$'(- spin flip probability) to denote the probability of an electron in a spin up edge mode to scatter into spin down edge mode or vice versa. If $f=0$ then there is no spin flip scattering betwixt the quasi-helical QSH edge modes and they behave as topological helical QSH edge modes. For $f>0$ they behave as quasi-helical QSH edge modes with scattering `$f$', while $f=1$ describes maximum scattering between them which reduces the transmission of electrons through the quasi-helical QSH edge modes to zero. As before, two of the probes are considered to be disordered, i.e., transmission probability through those contacts are $0<T_i<1$, $i=1,4$. Currents and voltages at the various contacts can be calculated following Eqs.~(2,3,4) with $N_\sigma$ being the number of edge modes with spin $\sigma=\uparrow/\downarrow$ electron ( scattering matrix elements $s^{\sigma\sigma'}_{ij}$ are explicitly calculated in the Appendix section B),
\begin{eqnarray}
I=GV
\end{eqnarray}
with
\begin{equation}
G=\frac{e^{2} }{h} \left( \begin{array}{cccccc}
    T_{11}  & -T_{12} & -T_{13} & -T_{14}  & -T_{15} & -T_{16}\\
    -T_{21}  & T_{22} & -T_{23} & -T_{24}& -T_{25} & -T_{26} \\ 
    -T_{31}  & -T_{32} & T_{33} & -T_{34}& -T_{35} & -T_{36} \\
     -T_{41}  & -T_{42} & -T_{43} & T_{44} & -T_{45} & -T_{46}\\ 
     -T_{51}  & -T_{52} & -T_{53} & -T_{54}& T_{55} & -T_{56} \\
     -T_{61}  & -T_{62} & -T_{63} & -T_{64} & -T_{65} & T_{66}\\ 
     \end{array} \right)
\end{equation}
where, the transmission probabilities $T_{ij}$'s are calculated in the Appendix section B. Choosing  reference potential at contact $V_{4}=0$, and since $I_2=I_3=I_5=I_6=0$ (these are voltage probes), we get the longitudinal and Hall resistance for the QAH+ case as $R_L^{QAH+}=R_{23,14}$ and $R_H^{QAH+}=R_{26,14}$ respectively. For ideal case, i.e., the strength of disorder $D_1=D_4=0$, these expressions for $R_L^{QAH+}$ and $R_H^{QAH+}$ reduce to $\frac{h}{e^2}\frac{2-3f+f^2}{9-15f+9f^2-2f^3}$ and $\frac{h}{e^2}\frac{3-2f}{9-15f+9f^2-2f^3}$. As the expressions in general for finite disorder for both $R^{QAH+}_{L}$ and $R^{QAH+}_{H}$ are quite large, so we plot them in Fig.~3(a,b). Increase in disorder implies increase in elastic scattering which in general occurs when temperature increases. In Fig.~3(a,b), we see for the QAH case (single chiral QAH edge mode) both the Hall $R_H^{QAH}$ as well the longitudinal resistance $R_L^{QAH}$ are independent of disorder but for the QAH+ case (single chiral QAH edge mode exists with quasi-helical QSH edge modes) both $R_L^{QAH+}$ as well as $R_{H}^{QAH+}$ are dependent on disorder and this dependence mirrors that seen in Fig.~1(a,b) which was also observed in Refs.~\cite{bestwick, cui}. To conclude this subsection the role of temperature in any mesoscopic experiment is to increase the amount of electron-phonon scattering as temperature increases, which in general leads to inelastic scattering. Therefore for a proper comparison with the experiments of Refs.~\cite{bestwick, cui} we need to include inelastic scattering also. {Here, of course the inelastic scattering is included phenomenologically and is at zero temperature too implying that it is a model for electron-electron interaction rather than electron-phonon interaction.}

\subsubsection{Why we do not consider all probe disorder}
We can calculate for QAH case the conductance in presence of all disordered contacts. However, for QAH+ case, as we increase the number of disordered
contacts in our problem, it increases the difficulty in calculating the transmission
probabilities due to increase in number of trajectories
between different contacts. If we consider all the contacts to be disordered then
to calculate the transmission probabilities would be
almost impossible for QAH+ case. Thats why we have chosen only two contacts
1 and 4 are to be disordered which are widely separated for simplicity only.

\subsection{ Disordered 6T QAH+ bar with inelastic scattering} The effect of both disorder as well as inelastic scattering is taken into consideration now, see Fig.~2(d). The topological QAH edge mode equilibrates to a new potential $V''_i$ ($i=1-6$) and does not equilibrate its energy with the quasi-helical QSH edge modes. On the other hand quasi-helical QSH edge modes equilibrate between themselves to a new potential $V'_i$. The spin flip probability between the QSH (quasi-helical) edge modes is $f$. The probability of an electron coming out of contact $1$ via up spin polarized QAH (topological) edge mode is $T_1$. To calculate the probability of an electron coming out of terminal $1$ via spin up/down quasi-helical QSH edge modes, we need to consider the scattering amplitudes, as due to the presence of disorder along with spin flip scattering between the quasi-helical QSH edge modes there are multiple paths from one contact to another. The scattering amplitude $s^{\uparrow\uparrow}_{11}$, the reflection amplitude from contact $1$ to itself with initial spin up to final spin up via quasi-helical QSH edge modes is $r_1-\frac{t_1^2r_1f}{a}$ with $a=1-R_1f$. This derivation is shown in Appendix section C. The transmission probability $T^{\uparrow\uparrow}_{11}$ for the quasi-helical QSH edge mode can be derived following Eq.~(3), which is $1-(R_1 + T_1^2 R_1 f^2/a^2 - 2 R_1 T_1 f/a)$. The scattering amplitude $s^{\downarrow\uparrow}_{11}$ for the quasi-helical QSH edge mode is $\frac{t_1^2\sqrt{f}}{a}$, which implies transmission probability $T^{\downarrow\uparrow}_{11}=\frac{T_1^2f}{a}$. Similarly, we can see that $T^{\downarrow\downarrow}_{11}=T^{\uparrow\uparrow}_{11}$ and $T^{\uparrow\downarrow}_{11}=T^{\downarrow\uparrow}_{11}$. So, the total probability of an electron to come out of contact $1$ via the quasi-helical QSH edge mode is $T_{11}=\sum_{\sigma,\sigma'}T^{\sigma\sigma'}_{11}=(2- 2 (R_1 + T_1^2 R_1 f^2/a^2 + T_1^2 f/a^2 - 2 R_1 T_1 f/a))$. We need to add this transmission probability via the quasi-helical QSH edge mode with that for the chiral QAH edge mode, by doing so we get the total probability $T_{11}=T_1(3 - 2 f + f^2 (-2 + R_1) R_1)/a^2$. So the current coming out of contact $1$ is $I_1^{out}=\frac{e^2}{h}T_1(3 - 2 f + f^2 (-2 + R_1) R_1)/a^2V_1$. The currents coming into contact $1$ are from equilibrating potentials $V''_6$, $V'_1$ and $V'_6$. The currents coming into contact $1$ via the topological QAH edge mode is $\frac{e^2}{h}T_1V_6''$, while via quasi-helical QSH edge modes are $\frac{e^2}{h}(T_1(1-f)/a^2+T_1R_1(1-f)f/a^2)(V_1'+V_6')$ (explained in the Appendix section C). Thus, $I_1^{in}=\frac{e^2}{h}(T_1(1-f)/a^2+T_1R_1(1-f)f/a^2)(V_1'+V_6')+\frac{e^2}{h}T_1V_6''$. The net current out of contact $1$ is then
\begin{eqnarray}
I_1&=&I_1^{out}-I_1^{in}=\frac{e^2}{h}[(T_1(3 - 2 f + f^2 (-2 + R_1) R_1)/a^2)V_1\nonumber\\&-&T_1V''_{1}-(T_1(1-f)/a^2+R_1 T_1 (1 - f) f/a^2)(V'_1+V'_{6})]\nonumber\\ \qquad\text{with }& & I_4=-I_1, \text{where $a=1-R_1f$ and $c=1-R_4f$.}
\end{eqnarray}
 The currents at rest of the contacts (explained in the Appendix section C) are as follows-
\begin{equation}
I_i=\frac{e^2}{h}[(3-2f)V_i-(1-f)(V'_i+V'_{i-1})-V''_{i-1}], i=2,3,5,6.
\end{equation}
As stated above, equilibration in this case happens separately for QAH edge modes. We explain the equilibration at $V_1''$ below. Equilibration at the primes and double primed potentials can be derived likewise. The current coming out of contact $1$, $\frac{e^2}{h}T_1V_1$, via the chiral QAH edge mode is equilibrated with the current coming from equilibrating potential $V''_6$ $\frac{e^2}{h}R_1V''_6$ via equilibrating potential $V''_1$. So, the current coming into the equilibrating potential $V''_1$ is $\frac{e^2}{h}(T_1V_1+R_1V''_6)$ and the current coming out of that equilibrating potential is $\frac{e^2}{h}V''_1$. This two currents are equilibrated, i.e., $\frac{e^2}{h}(T_1V_1+R_1V''_6)=\frac{e^2}{h}V''_1$.  Similarly, other potentials $V''_i$ are related to $V_i$ by-
\begin{eqnarray}
V''_2=V_2, \quad V''_5=V_5, \quad R_1V_6''+T_1V_1=V''_1\nonumber\\
V''_3=V_3, \quad V''_6=V_6, \quad R_4V''_3+T_4V_4=V''_4
\end{eqnarray}
and the relations between potentials $V'_{i}$ (after equilibration) and $V_{i}$ are stated in the Appendix section C.
Choosing potential at reference contact 4- $V_{4}=0$, and since contacts 2, 3, 5 and 6 are voltage probes we have: $I_2=I_3=I_5=I_6=0$. We thus derive longitudinal resistance $R_L^{QAH+}=\frac{h}{e^2}\frac{3-4f+f^2}{14-15f+6f^2-f^3}$ and $R_H^{QAH+}=\frac{h}{e^2}\frac{4}{7-4f+f^2}$ in absence of disorder but in presence of inelastic scattering. The expressions for both $R_L^{QAH+}$ as well as $R_H^{QAH+}$ in presence of disorder as well as inelastic scattering are quite sizable so we scrutinize them via plots as in Figs.~4(a,b). In Figs.~4(a, b) we see that  when both disorder and inelastic scattering is taken into consideration, for a single chiral QAH edge mode the longitudinal resistance $R_L^{QAH}$ is zero while the Hall resistance $R_H^{QAH}$ is quantized as before implying that for a single a chiral QAH edge mode the transport coefficients are independent of disorder as well as inelastic scattering. Now further for a chiral QAH edge mode existing with quasi-helical QSH edge modes ($QAH+$ case) $R_H^{QAH+}$ loses its quantization while $R_L^{QAH+}$ increases with disorder showing a similar behavior as was observed in Refs.~\cite{bestwick, cui} with temperature and also shown in Figs.~1(a,b). {The data in Fig.~1 shows how a finite longitudinal resistance ($R_L$) and deviation of Hall resistance ($R_H$) from the quantized value arise as temperature increases. Our results show that a finite $R_L$ and non-quantized $R_H$ arise with disorder even at zero temperature. We have shown tha the effect of temperature on edge mode transport is similar to the effect of scattering via disorder. As temperature increases the scattering of electrons increases due to the decrease in elastic scattering length $l_e$ and also of phase coherence length $l_\phi$. If $l_e$ decreases the scattering of electrons increases via disorder and similarly if $l_\phi$ decreases then scattering of electrons increases due to inelastic scattering. In our paper, as we increase the disorder strength from zero (no disorder) to maximum (complete disorder), independent of inelastic scattering, we see that longitudinal resistance increases exponentially while the Hall resistance deviates from its quantized value. Inelastic scattering has a weak temperature dependence as shown in Fig.~4. The finite $R_L$ and non-quantized $R_H$ are mainly due to presence of disordered contacts. Inelastic scattering does not lead to any qualitative change only the magnitude of $R_L$ and $R_H$ changes.}
\begin{figure}
\centering
{\includegraphics[width=0.5\textwidth]{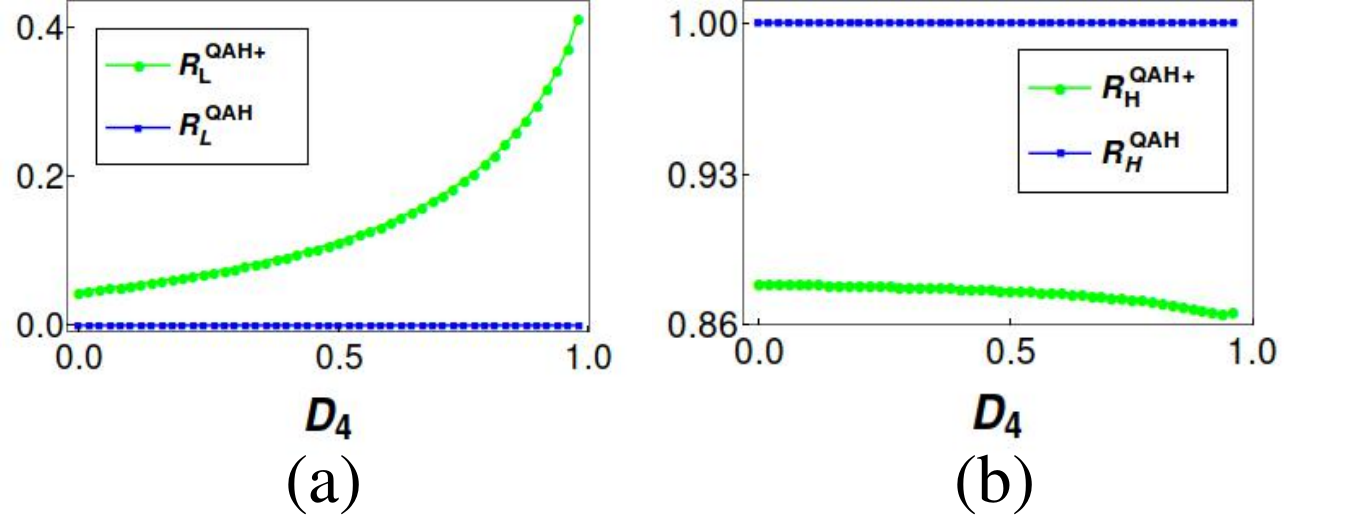}}
\caption{ (a) The longitudinal resistance and (b) the Hall resistance (both in units of $e^2/h$) vs disorder ($D_4$) at terminal $4$ in presence of inelastic scattering is shown for a 6 terminal QAH sample. Parameters are $D_1=0.5$, spin-flip scattering $f=0.9$.}
\end{figure}

\section{Conclusion} We see that presence of finite longitudinal resistance and deviation from Hall resistance quantization in a QAH insulator exists not only in presence of temperature as explained in Refs.~\cite{bestwick, cui} but also due to the presence of quasi-helical helical QSH edge modes along with chiral QAH edge mode. We can conclude that experiments as shown in Ref.~\cite{bestwick, cui} can not just be interpreted as indication of chiral QAH effect but rather of QAH edge mode occurring with quasi-helical QSH edge modes and further temperature is not the sole reason for seeing finite $R_L$, existence of quasi-helical edge modes could also be a plausible reason. 
 \acknowledgements This work was supported by funds from Dept. of Science and Technology (SERB), Govt. of India, Grant No. EMR/2015/001836.
  
 \section{Appendix}
 The scattering matrix  for disordered QAH and QAH+ bar are derived. Further the process of equilibration of currents for disorder with inelastic scattering in QAH+ bar are dealt with.

{\subsection{Disordered 6T QAH bar}} For chiral (topological) QAH edge mode, the scattering matrix $S$ in presence of disorder is shown below (see Fig.~2(a))-
\begin{equation}
S= \left( \begin{array}{cccccc}
   r_1 & 0 & 0 & 0&0&t_1 \\
    -t_1 & 0 & 0 & 0&0&r_1\\ 
   0  & 1 & 0& 0&0&0 \\
     0  & 0 & t_4&  r_4&0&0 \\
      0  & 0 & r_4 &  -t_4&0&0 \\
      0  & 0 & 0&  0&1&0 \\
      \end{array} \right),
\end{equation}
where, $r_i$ and $t_i$ are the reflection and transmission amplitude at the contact `$i$'. The transmission probabilities $T_{ij}$'s can be easily calculated by following $T_{ij}=Tr[s^\dagger_{ij}s_{ij}]$ (when $i\neq j$) and $T_{ii}=1-Tr[s^\dagger_{ii}s_{ii}]$ relations (Eq.~(5) in the main manuscript is derived from this scattering matrix). Since here the edge mode consists of only spin up electrons, and no spin flip scattering, we have omitted the spin indexes $\sigma,\sigma'$ from the S matrix elements. \\
 \subsection{ Disordered 6T QAH+ bar}
 For chiral(topological) QAH edge mode with quasi-helical QSH edge modes the scattering matrix is a $6\times6$ square matrix, pertaining to the six terminals of the QAH bar. However, since there are three edge modes one spin up QAH and spin up and spin down QSH edge modes, arising from each contact,
 \begin{equation}
S= \left( \begin{array}{cccccc}
  s_{11}  & s_{12} & s_{13} & s_{14}  & s_{15} & s_{16}\\
    s_{21}  & s_{22} & s_{23} & s_{24}& s_{25} & s_{26} \\ 
    s_{31}  & s_{32} &s_{33} & s_{34}& s_{35} & s_{36} \\
     s_{41}  & s_{42} & s_{43} & s_{44} & s_{45} & s_{46}\\ 
     s_{51}  & s_{52} & s_{53} & s_{54}& s_{55} & s_{56} \\
     s_{61}  & s_{62} & s_{63} & s_{64} & s_{65} &s_{66}\\ 
     \end{array} \right)
\end{equation}
each element $s_{ij}$ of the $S$ matrix is a $3\times3$ matrix, as shown below-
 \begin{equation}
s_{ij}=
  \left( \begin{array}{ccc}
  \chi^{\uparrow\uparrow}  &0& 0\\
    0  & \tau^{\uparrow\uparrow} & \tau^{\uparrow\downarrow}\\ 
    0  & \tau^{\downarrow\uparrow}&\tau^{\downarrow\downarrow}\\
     \end{array} \right),
\end{equation}
in the basis $(\uparrow_{QAH},\uparrow_{QSH},\downarrow_{QSH})^T$. The $(1,1)$ component of these $s_{ij}$ matrices depicts chiral (topological) QAH edge mode, while $(2,2), (2,3), (3,2), (3,3)$ elements correspond to the scattering between different spin components of quasi-helical QSH edge modes while $(1,2), (1,3), (2,1), (3,1)$ which imply scattering between chiral (topological) QAH edge mode and quasi-helical QSH edge modes are all zero as there is no scattering between QAH and two QSH edge modes. Writing all the $s_{ij}$ matrices in Eq.~(16), we have the following 14 zero matrices:
\begin{eqnarray}
s_{13}=s_{31}&=&s_{14}=s_{41}=s_{42}=s_{42}=s_{15}=\bar{0},\nonumber\\s_{51}=s_{25}&=&s_{52}=s_{36}=s_{63}=s_{48}=s_{64}=\bar{0},\\
\text{with }\bar{0}&=& \left( \begin{array}{ccc}
 0  &0& 0\\
    0  & 0& 0\\ 
    0  & 0&0\\
     \end{array} \right),\nonumber
\end{eqnarray}
and rest of the matrices are as follows-
\begin{widetext}
{\small
\begin{eqnarray}
s_{11}&=& \left( \begin{array}{ccc}
 r_1 &0& 0\\0  &\frac{r_1(1-f)}{a}&-\frac{t_1^2\sqrt{f}}{a}\\0  &-\frac{t_1^2\sqrt{f}}{a} &\frac{r_1(1-f)}{a} \end{array} \right),
s_{12}= \left( \begin{array}{ccc}
0&0&0\\0&\frac{t_1\sqrt{1-f}}{a}&\frac{t_1r_1\sqrt{(1-f)f}}{a}\\0&0&0 \end{array} \right),
s_{16}= \left( \begin{array}{ccc}
t_1&0&0\\0&0&0\\0&\frac{t_1r_1\sqrt{(1-f)f}}{a}&\frac{t_1\sqrt{1-f}}{a} \end{array} \right),
s_{21}= \left( \begin{array}{ccc}
 - t_1&0&0\\0&0&-\frac{t_1r_1\sqrt{(1-f)f}}{a}\\ 0&0&-\frac{t_1\sqrt{1-f}}{a}\end{array} \right),\nonumber\\
 s_{22}&=& \left( \begin{array}{ccc}
 0&0&0\\0&0&-\frac{\sqrt{f}t_1^2}{a}\\0&\sqrt{f}&0\end{array} \right),
 s_{23}= \left( \begin{array}{ccc}
 0&0&0\\0&\sqrt{1-f}&0\\0&0&0\end{array} \right),
 s_{26}= \left( \begin{array}{ccc}
 r_1&0&0\\0&0&0\\0&0&\frac{r_1(1-f)}{a}\end{array} \right),
 s_{32}= \left( \begin{array}{ccc}
 1&0&0\\0&0&0\\0&0&-\sqrt{1-f}\end{array} \right),
 s_{33}= \left( \begin{array}{ccc}
 0&0&0\\0&0&\sqrt{f}\\0&-\frac{\sqrt{f}t_4^2}{c}&0\end{array} \right),\nonumber\\
 s_{34}&=& \left( \begin{array}{ccc}
 0&0&0\\0&-\frac{t_4\sqrt{1-f}}{c}&0\\0&-\frac{t_4r_4\sqrt{(1-f)f}}{c}&0\end{array} \right),
 s_{35}= \left( \begin{array}{ccc}
 0&0&0\\0&\frac{r_4(1-f)}{c}&0\\0&0&0\end{array} \right),
  s_{43}= \left( \begin{array}{ccc}
t_4&0&0\\0&0&0\\0&\frac{t_4r_4\sqrt{(1-f)f}}{c}&\frac{t_4\sqrt{1-f}}{c}\end{array} \right),
 s_{44}= \left( \begin{array}{ccc}
 r_4&0&0\\0&\frac{r_4(1-f)}{c}&-\frac{t_4^2\sqrt{f}}{c}\\0&-\frac{t_4^2\sqrt{f}}{c}&\frac{r_4(1-f)}{c}\end{array} \right),\nonumber\\
 s_{45}&=& \left( \begin{array}{ccc}
 0&0&0\\0&\frac{t_4\sqrt{1-f}}{c}&\frac{t_4r_4\sqrt{(1-f)f}}{c}\\0&0&0\end{array} \right),
  s_{53}= \left( \begin{array}{ccc}
  r_4&0&0\\0&0&0\\0&0&\frac{r_4(1-f)}{c}\end{array} \right),
  s_{54}= \left( \begin{array}{ccc}
  -t_4&0&0\\0&0&-\frac{t_4r_4\sqrt{(1-f)f}}{c}\\0&0&-\frac{t_4\sqrt{1-f}}{c}\end{array} \right),
  s_{55}= \left( \begin{array}{ccc}
  0&0&0\\0&0&-\frac{\sqrt{f}t_4^2}{c}\\0&\sqrt{f}&0\end{array} \right),\nonumber\\
   s_{56}&=& \left( \begin{array}{ccc}
   0&0&0\\0&\sqrt{1-f}&0\\0&0&0\end{array} \right),
   s_{61}= \left( \begin{array}{ccc}
   0&0&0\\0&-\frac{t_1\sqrt{1-f}}{a}&0\\0&-\frac{t_1r_1\sqrt{(1-f)f}}{a}&0\end{array} \right),
   s_{62}= \left( \begin{array}{ccc} 
   0&0&0\\0&\frac{r_1(1-f)}{a}&0\\0&0&0\end{array} \right),
    s_{65}= \left( \begin{array}{ccc} 
   1&0&0\\0&0&0\\0&0&-\sqrt{1-f}\end{array} \right),
   s_{66}= \left( \begin{array}{ccc} 
   0&0&0\\0&0&\sqrt{f}\\0&-\frac{\sqrt{f}t_1^2}{a}&0\end{array} \right).\nonumber\\
\end{eqnarray}}
\end{widetext}
where $a=1-r_1^2f$, and $c=1-r_4^2f$. 
We have explained the origin of $s_{11}$ matrix below. The transmission probability $T_{11}$ derived out of this matrix follows. Each of the three rows of these $s_{ij}$ matrices corresponds to three edge modes- one spin up chiral (QAH) edge mode and two spin up/down quasi-helical QSH edge modes. The $(1,1)$ element of $s_{11}$ matrix represents the reflection amplitude of an electron from contact $1$ to itself after suffering some scattering by the disorder at contact $1$ via the QAH edge mode, which is $r_1$. The (2,2) element of $s_{11}$ matrix represents the reflection of an spin up electron from contact $1$ to itself without any spin flip scattering via the QSH edge mode, which is $r_1-\frac{t_1^2r_1f}{1-r_1^2f}=\frac{r_1(1-f)}{1-r_1^2f}$. This can be explained in this way, a spin up electron can directly reflect from contact $1$ with the same spin and scattering amplitude $r_1$, but this one of the paths. It can also reflect after some scattering following a second path with amplitude $t_1^2r_1f$ or even a third path with amplitude $t_1^2r_1^3f^3$ and likewise fourth ... nth path also. When we sum over all the paths we get the total scattering amplitude $s^{\uparrow\uparrow}_{11}=\frac{r_1(1-f)}{1-r_1^2f}$. Similarly, the scattering amplitude $s^{\uparrow\downarrow}_{11}$ consists of multiple paths from contact $1$ to itself from initial spin down to final spin up-which is summed up as $\frac{t_1^2\sqrt{f}}{1-r_1^2f}$. Here, whenever an electron emerges of a disordered contact, we put a phase shift $e^{i\pi}$, which is same as having a phase $\pi/2$ for an electron to enter or leave the contact, which leads to the unitarity of the total scattering matrix $S$ of the system.
The matrix $s_{11}$ thus is written below-
 \begin{equation}
s_{11}= \left( \begin{array}{ccc}
   r_1 &0& 0\\
    0  &\frac{r_1(1-f)}{1-r_1^2f}&-\frac{t_1^2\sqrt{f}}{1-r_1^2f}\\ 
    0  &-\frac{t_1^2\sqrt{f}}{1-r_1^2f} &\frac{r_1(1-f)}{1-r_1^2f}\\
     \end{array} \right).
\end{equation}
Thus, $s^{\uparrow\uparrow}_{11}=\left(\begin{array}{cc}\chi^{\uparrow\uparrow}&0\\0&\tau^{\uparrow\uparrow}\end{array}\right)=\left(\begin{array}{cc}r_1&0\\0&\frac{r_1(1-f)}{1-r_1^2f}\end{array}\right)$, $s^{\uparrow\downarrow}_{11}=s^{\downarrow\uparrow}_{11}=-\frac{t_1^2\sqrt{f}}{1-r_1^2f}$ and $s^{\downarrow\downarrow}_{11}=\frac{r_1(1-f)}{1-r_1^2f}$. The transmission probability $T^{\uparrow\uparrow}_{11}$ can be easily calculated by following: $T^{\uparrow\uparrow}_{11}=(N_\uparrow-Tr[s^{\uparrow\uparrow\dagger}_{11}s^{\uparrow\uparrow}_{11}])$, from Eq. (2), which is $2-((R_1 (2 - 2 f (1 + R_1) + f^2 (1 + R_1^2)))/(1 - f R_1)^2)$ (here $N_\uparrow=2,N_\downarrow=1$). Similarly, one can calculate $T^{\uparrow\downarrow}_{11}=T^{\downarrow\uparrow}_{11}=\frac{T_1^2f}{(1-R_1f)^2}$, and $T^{\downarrow\downarrow}_{11}=\frac{R_1(1-f)^2}{(1-R_1f)^2}$ to derive $T_{11}=\sum_{\sigma\sigma'}T_{11}=T_1(3 - 2 f + f^2 (-2 + R_1) R_1)/(1 - f R_1)^2$. All the other $s_{ij}$ matrices and corresponding $T_{ij}$ transmission probabilities can thus be derived, with
where, 
\begin{eqnarray}
T_{11}&=&T_1(3 - 2 f + f^2 (-2 + R_1) R_1)/a^2,\nonumber\\
T_{12}&=&T_1(1-f)(1 + f R1)/a^2,\nonumber\\
T_{16}&=&T_{21}=T_1(2 -f^2 T_1R_1 - f (1 + R_1))/a^2,\nonumber\\
T_{13}&=&T^{14}=T^{15}=0,\nonumber\\
T_{22}&=&(3 - f^3 R_1^2 + f^2 R_1 (2 + 3 R_1) - f (2 + 4 R_1 + R_1^2))/a^2,\nonumber\\
T_{23}&=&(1 - f),\nonumber\\
T_{26}&=&R_1 + R_1 (1 - f)^2/a^2,\nonumber\\
T_{25}&=&T_{24}=0,
\end{eqnarray}
with $a = 1 - R_1 f, c = 1 - R_4 f$. Replacing $R_1$ with $R_4$ in the above equation rest of the transmission probabilities $T_{31}$ to $T_{66}$ can be written.\\

\subsection{Disordered 6T QAH+ bar with inelastic scattering}
For chiral(topological) QAH edge mode with quasi-helical QSH edge modes the currents between the contacts are derived below. The current coming out of contact $1$ is $I_1^{out}=\frac{e^2}{h}T_1(3 - 2 f + f^2 (-2 + R_1) R_1)/a^2V_1$, and the detailed derivation of $I_1^{out}$ has been given in the main manuscript. The incoming currents into contact $1$ from equilibrating potentials $V''_6$, $V'_1$ and $V'_6$ is derived below. The current coming into contact $1$ via the topological QAH edge mode is $\frac{e^2}{h}T_1V_6''$. To calculate the current coming into contact $1$ from equilibrating potentials $V'_1$ and $V'_6$ via the quasi-helical QSH edge modes, we look at the scattering matrix amplitudes. The scattering amplitude $s^{\downarrow\downarrow}_{11'}$ of an electron to enter contact $1$ from equilibrating potential $V_1'$ via the spin down quasi-helical QSH edge mode is $t_1\sqrt{1-f}$, but this is the one of the paths, it can also enter the contact $1$ by following a second path with scattering amplitude $t_1r_1^2f\sqrt{1-f}$ or a third path with scattering amplitude $t_1r_1^4f^2\sqrt{1-f}$ and so on for infinite no. of paths. After summing over all the scattering paths we get the total scattering amplitude $t_1\sqrt{1-f}/a$ with the transmission probability $T^{\downarrow\downarrow}_{11'}=T_1(1-f)/a^2$. Similarly, one can derive the scattering amplitude $s^{\uparrow\downarrow}_{11'}=t_1r_1\sqrt{f(1-f)}/a$, giving the transmission probability $T^{\uparrow\downarrow}_{11'}=T_1R_1(1-f)f/a^2$. Thus, the current coming into contact $1$ from equilibrating potential $V_1'$ is $\frac{e^2}{h}(T_1(1-f)/a^2+T_1R_1(1-f)f/a^2)V_1'$. The probability of an electron coming to contact $1$ from equilibrating potential $V_6'$ is same as that coming from $V_1'$, this can be seen from Fig.~2(d). So, the current coming into contact 1 from equilibrating potential $V_6'$ is $\frac{e^2}{h}(T_1(1-f)/a^2+T_1R_1(1-f)f/a^2)V_6'$, thus $I_1^{in}=\frac{e^2}{h}(T_1(1-f)/a^2+T_1R_1(1-f)f/a^2)(V_6'+V_1')+\frac{e^2}{h}T_1V_1''$. Thus, the net current out of contact $1$ is- 
\begin{eqnarray}
I_1&=&=I_1^{out}-I_1^{in}=\frac{e^2}{h}[(T_1(3 - 2 f + f^2 (-2 + R_1) R_1)/a^2)V_1\nonumber\\&-&T_1V''_{1}-(T_1(1-f)/a^2+R_1 T_1 (1 - f) f/a^2)(V'_1+V'_{6})]\nonumber\\& & \qquad\text{with } I_4=-I_1.
\end{eqnarray}
The current at other contacts $2, 3, 5, 6$ can be calculated easily by considering the transmission probabilities alone as these contacts are not affected by any disorder. Similarly, current passing through contact $2$ can be derived as follows. The current coming out of contact $2$ via single QAH edge mode is $\frac{e^2}{h}V_2$, and via spin up and spin down quasi-helical QSH edge modes is $\frac{e^2}{h}2(1-f)V_2$. Thus the total current coming from contact $2$ is $I_2^{out}=\frac{e^2}{h}(3-2f)V_2$, while currents coming into contact $2$ from equilibrating potentials $V''_1$, $V'_1$ and $V_2'$ via single QAH edge mode is $\frac{e^2}{h}V''_1$, and via spin up and down quasi-helical QSH edge modes are $\frac{e^2}{h}(1-f)(V'_2+V'_{1})$ respectively, thus $I_2^{in}=\frac{e^2}{h}(1-f)(V'_2+V'_{1})+\frac{e^2}{h}V''_1$. So the net current through contact $2$ is $I_2=I_2^{out}-I_2{in}=\frac{e^2}{h}[(3-2f)V_2-(1-f)(V'_1+V'_{2})-V''_{1}]$.
 The currents at rest of the contacts can be similarly derived-
\begin{eqnarray}
I_i&=&\frac{e^2}{h}[(3-2f)V_i-(1-f)(V'_i+V'_{i-1})-V''_{i-1}], \text{ for } i=2,3,5,6.\nonumber\\
\end{eqnarray}
with $a=1-R_1f$ and $c=1-R_4f$. Next we equilibrate the currents at the primed potentials as follows. The current coming into the equilibration potentials $V'_2$ from contact $2$ is $\frac{e^2}{h}(1-f)V_2$ and from contact $3$ is $\frac{e^2}{h}(1-f)V_3$, while the current coming out of $V'_2$ after equilibration is $\frac{e^2}{h}2(1-f)V'_2$. These incoming and outgoing currents at equilibrating potential $V_2'$ must be equal, thus $\frac{e^2}{h}(1-f)V_2+\frac{e^2}{h}(1-f)V_3=\frac{e^2}{h}2(1-f)V_2'$. Similar relations can be derived for the other primed potentials $V_3'-V_6'$, $V_1'$ and they are mentioned below:
{\small
\begin{eqnarray}
(1-f)(V_i+V_{i+1})&=&2(1-f)V'_i, \quad\text{for }i=2, 5\nonumber\\
(1-f)V_{i}+(\frac{T_{i+1}(1-f)}{1-R_{i+1} f}&+&\frac{T_{i+1}R_{i+1}f(1-f)}{1-R_{i+1}f})V_{i+1}+\frac{R_i(1-f)^2}{1-R_{i+1}f}V'_{i+1}\nonumber\\=((1-f)+\frac{T_{i+1}(1-f)}{1-R_{i+1}f}&+&\frac{T_{i+1}R_{i+1}f(1-f)}{1-R_{i+1}f}+\frac{R_{i+1}(1-f)^2}{1-R_{i+1}f})V'_i,\nonumber\\\quad\text{for }i=3,6, \text{ and},\nonumber\\
(1-f)V_{i+1}+(\frac{T_{i}(1-f)}{1-R_{i} f}&+&\frac{T_{i}R_{i}f(1-f)}{1-R_{i}f})V_{i}+\frac{R_i(1-f)^2}{1-R_if}V'_{i-1}\nonumber\\=((1-f)+\frac{T_{i}(1-f)}{1-R_{i}f}&+&\frac{T_{i}R_{i}f(1-f)}{1-R_{i}f}+\frac{R_{i}(1-f)^2}{1-R_{i}f})V'_i,\nonumber\\\quad\text{for }i=1,4.
\end{eqnarray}
}


\begin{thebibliography}{99}
  \bibitem{bestwick} A. J. Bestwick, et. al., Precise Quantization of the Anomalous Hall Effect near Zero Magnetic Field, Phys. Rev. Lett. 114, 187201 (2015).
  \bibitem{klitzing} K. V. Klitzing, et. al., New Method for High-Accuracy Determination of the Fine-Structure Constant Based on Quantized Hall Resistance, Phys. Rev. Lett. 45, 494 (1980).
  \bibitem{roth} A. Roth, et. al., Nonlocal Transport in the Quantum Spin Hall State, Science 325, 294 (2009).
   \bibitem{dong} D. Tong, Lectures on the Quantum Hall Effect, arXiv:1606.06687.
  \bibitem{cui} C.-Z. Chang, et. al., High-precision realization of robust quantum anomalous Hall state in a hard ferromagnetic topological insulator, Nature Mat. 14, 473-477 (2015).
\bibitem{kou} X. Kou, et. al., Scale-Invariant Quantum Anomalous Hall Effect in Magnetic Topological Insulators beyond the Two-dimensional Limit, Phys. Rev. Lett. 113, 137201 (2014).
\bibitem{che} J. G. Checkelsky, et. al., Trajectory of Anomalous Hall Effect toward the Quantized State in a Ferromagnetic Topological Insulator, Nature Phys. 10, 731-736 (2014).
 \bibitem{wang1} J. Wang, et. al., Anomalous Edge Transport in the Quantum Anomalous Hall State, Phys. Rev. Lett. 111, 086803 (2013).
 \bibitem{wang} J. Wang, et. al., Quantum anomalous Hall effect in Magnetic topological insulators, Phys. Scr. T164, 014003 (2015). 
\bibitem{sanvito} A. Narayan \& S. Sanvito, Multiprobe Quantum Spin Hall Bars, Eur. Phys. J. B 87: 43 (2014).
 \bibitem{chulkov} A. P. Protogenov, et. al., Nonlocal Edge State Transport in Topological Insulators, Phys. Rev. B 88, 195431 (2013).
\bibitem{Arjun} A. Mani \& C. Benjamin, Are quantum spin Hall edge modes more resilient to disorder, sample geometry and inelastic scattering than quantum Hall edge modes?, J. Phys.: Condens. Matter 28 (2016) 145303.
\bibitem{mani} A. Mani \& C. Benjamin, Fragility of non-local edge mode transport in the quantum spin Hall state, Phys. Rev. Applied 6, 014003 (2016).
{\bibitem{arjun3} A. Mani \& C. Benjamin, Role of helical edge modes in the
chiral quantum anomalous Hall
state, Scientific Reports 8: 1335 (2018).
\bibitem{mogi} M. Mogi, et. al., Magnetic modulation doping in topological insulators toward higher-temperature
quantum anomalous Hall effect, Appl. Phys. Lett. 107, 182401 (2015).
\bibitem{liu} Minhao Liu, et. al., Large discrete jumps observed in the transition
between Chern states in a ferromagnetic
topological insulator, Science Advances 2, e1600167 (2016). }
\bibitem{buti-sci} M. Buttiker, Edge-State Physics Without Magnetic Fields, Science 325, 278 (2009).
 \bibitem{buti} M. Buttiker, Absence of Backscattering in The Quantum Hall effect in Multiprobe Conductors, Phys. Rev. B 38, 9375 (1988); M. Buttiker, Transmission probabilities and the quantum Hall effect, Surface Science 229, 201 (1990).
 \bibitem{jen} Jens Nikolajsen, Bachelors project, Edge States and Contacts in
the Quantum Hall Effect, Nanoscience center, Niels Bohr Institute, Faculty of Science, Univ. of Copenhagen, Denmark.
 \bibitem{yan} Yanxia Xing, et. al., Influence of dephasing on the quantum Hall effect and the spin Hall effect, Phys. Rev. B 77, 115346(2008).
 \bibitem{rok} Roksana Golizadeh-Mojarad and Supriyo Datta, Nonequilibrium Green's function based models for dephasing in quantum transport, Phys. Rev. B 75, 081301(R) (2007).
 \bibitem{nikolajsen}  J. Nikolajsen, Bachelors Thesis, Edge States and Contacts in the Quantum Hall Effect, Nanoscience center, Niels Bohr Institute, Faculty of Science,  Univ. of Copenhagen, Denmark. 
 \bibitem{arjun} A. Mani \& C. Benjamin, Probing helicity and the topological origins of helicity via non-local Hanbury-Brown and Twiss correlations, Scientific Reports 7: 6954 (2017).
 \end{thebibliography}
\end{document}